\documentclass[sn-apacite]{sn-jnl}

\usepackage{graphicx}
\usepackage{caption}
\usepackage{subcaption}
\usepackage{hyperref}
\usepackage{amsmath}
\usepackage{afterpage}
\usepackage{bm}
\graphicspath{ {figs/} }

\bibpunct[, ]{(}{)}{;}{a}{,}{,}

\theoremstyle{thmstyleone}%
\theoremstyle{thmstyletwo}%

\theoremstyle{thmstylethree}%

\begin{document}

\title[IMaSC - ICFOSS Malayalam Speech Corpus]{IMaSC - ICFOSS Malayalam Speech Corpus}


\author*[1,4]{\fnm{Deepa} \sur{P Gopinath}}\email{deepa.gopinath@gecbh.ac.in}

\author*[2]{\fnm{Thennal} \sur{D K}}\email{thennal21bcs14@iiitkottayam.ac.in}

\author[3,4]{\fnm{Vrinda} \sur{V Nair}}\email{vrinda66nair@gmail.com}
\equalcont{These authors contributed equally to this work.}

\author[5]{\fnm{Swaraj} \sur{K S}}\email{swarajks10@gmail.com}
\equalcont{These authors contributed equally to this work.}

\author[6]{\fnm{Sachin} \sur{G }}\email{sachingracious@gmail.com}
\equalcont{These authors contributed equally to this work.}

\affil[1]{\orgdiv{Dept. of Electronics and Commn. Engg}, \orgname{Government Engineering College}, \orgaddress{\street{Kozhikode}, 
\state{Kerala}, \country{India}}}

\affil*[2]{\orgdiv{Dept.of Computer Science}, \orgname{IIIT}, \orgaddress{\street{Kottayam}, 
\state{Kerala}, \country{India}}}

\affil[3]{\orgdiv{} \orgname{State Project Facilitation Unit}, \orgaddress{\street{Trivandrum}, 
\state{Kerala}, \country{India}}}

\affil[4]{\orgdiv{A P J Abdul Kalam Technological University} \orgname{}\orgaddress{\street{},
\state{Kerala}, \country{India}}}

\affil[5]{\orgdiv{Product Manager}, \orgname{Trivandrum}\orgaddress{\street{},
\state{Kerala}, \country{India}}}

\affil[6]{\orgdiv{Software Developer}, \orgname{Trivandrum}\orgaddress{\street{},
\state{Kerala}, \country{India}}}

\abstract{Modern text-to-speech (TTS) systems use deep learning to synthesize speech increasingly approaching human quality, but they require a database of high quality audio-text sentence pairs for training. Malayalam, the official language of the Indian state of Kerala and spoken by 35+ million people, is a low resource language in terms of available corpora for TTS systems. In this paper, we present \emph{IMaSC}, a Malayalam text and speech corpora containing approximately 50 hours of recorded speech. With 8 speakers and a total of 34,473 text-audio pairs, \emph{IMaSC} is larger than every other publicly available alternative. We evaluated the database by using it to train TTS models for each speaker based on a modern deep learning architecture. Via subjective evaluation, we show that our models perform significantly better in terms of naturalness compared to previous studies and publicly available models, with an average mean opinion score of 4.50, indicating that the synthesized speech is close to human quality.}




\maketitle
\section{Introduction}\label{sec1}

Malayalam, one among the 22 scheduled languages\footnote{Languages included in the VIII schedule of the Constitution of India} of India, is the official language of the state of Kerala and union territories Lakshadweep and Puducherry. According to the Census of India (2011), Malayalam is the native language of around 35 million people in India. A South Dravidian subgroup of the Dravidian language family, it is also spoken by bilingual communities in contiguous parts of Karnataka and Tamil Nadu and by Malayali communities in various parts of the world (\url{https://www.britannica.com/topic/Malayalam-language})\footnote{Total users in all countries: 37,212,270 (L1: 36,512,270, L2: 700,000)}. Malayalam orthogrophy is phonemic, with a one-to-one mapping of graphemes and phonemes with very few exceptions \citep{manghat2020malayalam}.  Over the years, it has incorporated various aspects from other languages, with Sanskrit and later English being the most noteworthy examples \citep{bright1999re}. Other major languages whose vocabulary was integrated over the millennia include Arabic, Dutch, Hindustani, Pali, Persian, Portuguese, Prakrit, and Syriac \citep{skunjanpillai}.

Like many other Indian languages, Malayalam is also a low resourced language in terms of availability of text-to-speech corpora. Choudhary et al.  has noted that the information technology support in Indian languages has been lagging by decades compared to other languages like English, Japanese or Russian, because of several factors including a lack of sufficient language resources required for the development of such technology {\citep{choudhary2021ldc}}. 

 As an agglutinative language, complex words can be formed in Malayalam by combining smaller words and morphemes \citep{premjith2018deep}. In Malayalam, there is no absolute limit on the length and extent of agglutination. This results in a wide variation in the number of graphemes per word, and inflections at word or morpheme boundaries. These features engender the formation of a large vocabulary that necessitates compilation of large corpora for speech related applications \citep{srivastava2020indicspeech}.

In this paper we present the Malayalam text and speech corpus created for the development of TTS in Malayalam with the financial support of the International Centre for Free and Open Source Software (ICFOSS)\footnote{\url{https://icfoss.in/}}, an autonomous organization set up by the Government of Kerala for promoting free and open source software. Curated with the help of linguists, the corpus consists of 34,473 sentences and 49.63 hours of speech read in studio conditions by 8 speakers (4 male and 4 female).

  Our corpus, named \emph{IMaSC - ICFOSS Malayalam Speech Corpus}, was evaluated by building a TTS system for Malayalam based on a deep learning architecture. Deep learning is opted in this work since techniques based on it have been state-of-the-art in speech synthesis for many years \citep{srivastava2020indicspeech}. Deep learning algorithms learn from the given training data and build models for achieving the required functionality, and as such the database for training is critical in the performance of the model. In case of TTS, deep learning algorithms learn all parameters of the speech, the language and the speaker including intonation and duration patterns \citep{tan2021survey}. Deep learning systems are also known for its data hungry nature \citep{marcus2018deep}. These factors make a TTS system based on deep learning the best option to evaluate the sufficiency and quality of a speech corpus. \emph{IMaSC} is made publicly available via Kaggle\footnote{\url{https://kaggle.com/datasets/thennal/imasc}} and Hugging Face\footnote{\url{https://huggingface.co/datasets/thennal/IMaSC}}.

\section{Related works}

\subsection{TTS corpus in Indian Languages}

Owing to the lack of large text-to-speech corpora, the progress of developing reliable text-to-speech systems for Indian languages has been relatively slow \citep{srivastava2020indicspeech}. One of the first efforts in this area is reported by Prahallad et al. \citep{prahallad2012iiit}. Baby et al. later developed a much larger resource for Indian languages, IndicTTS, which contains about 8 hours of speech data for 13 Indian languages \citep{baby2016resources}. Pradahan et al. used this corpus to train text-to-speech systems for these 13 languages \citep{pradhan2015building}. However, the data provided for each language in IndicTTS is insufficient for training recent neural-network-based systems that can produce natural, accurate speech according to Srivastava et al., who presented IndicSpeech, a large scale text-to-speech corpus for 3 Indian languages aimed at training neural TTS systems \citep{srivastava2020indicspeech}. The mean opinion score obtained for the TTS model trained on Malayalam corpus is lower than those obtained for Hindi and Bengali corpora,  which they attribute to the fundamental characteristics of Malayalam, such as the morpho-phonemic changes during word formation. They suggest that one of the solutions would be to increase the size of the Malayalam corpus to cover a larger vocabulary. He et al. presented multi-speaker corpora for 6 Indian languages, with approximately 6 hours of data for Malayalam split between 42 speakers \citep{he-etal-2020-open}.

\subsection{TTS corpus evaluation using deep learning models}

A TTS speech corpus can be evaluated by testing the quality of synthetic speech generated with the corpus \citep{dybkjaer2007evaluation}. Ahmed et al. prepared a phonetically balanced Bangla corpus and evaluated it using a Bangla neural synthesizer based on Merlin, an open-source speech synthesis toolkit using deep neural networks \citep{ahmad2021sust}. Deep learning architectures based on Tacotron and Tacotron 2 were used for evaluation of CSS10 \citep{park2019css10}, LibriTTS \citep{zen2019libritts}, Latvian corpus created by Dar`gis et al. \citep{dargis2020development}, DiDiSpeech \citep{guo2021didispeech}, KazakhTTS \citep{mussakhojayeva2021kazakhtts}, and TTS-Portuguese Corpus \citep{casanova2022tts}. FastSpeech was used for the evaluation of AISHELL-3 \citep{shi2020aishell} and Didispeech \citep{guo2021didispeech}.

 Srivastava et al. released IndicSpeech, a corpus curated for 3 Indian languages---Hindi, Malayalam and Bengali---with Deep Voice 3 models trained to evaluate the corpus \citep{srivastava2020indicspeech}. BU-TTS, a bilingual Welsh-English speech corpus developed by Russell et al., was evaluated by training VITS models, instead of a two stage architecture comprising of an aligner training stage and a vocoder training stage like in most other architectures \citep{russell2022bu}.

\section{Method}
\subsection{Corpus design}\label{corpus}
 The task of compiling a phonetically rich corpus involves linguistic analysis of a large raw text corpus, which we did in collaboration with linguists from the Department of Linguistics, University of Kerala\footnote{\url{https://www.keralauniversity.ac.in/home}}.

\subsubsection{Text collection}
The text corpus was derived from Malayalam Wikipedia. Launched on December 21, 2002, the Malayalam edition is the leading Wikipedia among other South Asian language Wikipedias in various quality metrics (\url{https://ml.wikipedia.org/wiki}). It has grown to be a wiki containing 79,510 articles as of October 2022, and ranks 13th in terms of depth among Wikipedias (\url{https://en.wikipedia.org/wiki/Malayalam_Wikipedia}).
This choice was made primarily because the articles of Wikipedia are in public domain and the scale of the wiki is more than sufficient to compile a phonetically balanced database. This enabled us to select a set of phonetically balanced sentences, record the corresponding speech and release it in public domain without any copyright infringements.

\subsubsection{Preparation of text corpus}
A dump\footnote{Dump is a large amount of data moved from one computer system, file,
or device to another} of Malayalam Wikipedia was created by scraping\footnote{Web scraping is the method used for extracting data from websites} the website. Preliminary data cleaning was done on the text files created from the dump. The clean data thus obtained was separated into a set of sentences, and only sentences composed entirely of Malayalam characters and punctuation were kept. Sentences were sampled from this set and vetted by linguists for quality in terms of naturalness, semantics and syntax. The sentences selected were ensured to constitute a phonetically balanced text corpus.

\subsection{Speaker selection}\label{speakersel}
People capable of correct pronunciation, pleasing rhythm, and consistent articulation were engaged for recording speech for TTS. The details of the spectrum of speakers we engaged for our speech corpus generation are given in Table \ref{tab1}. People of varying professional experience were selected for achieving a diverse range of prosody and articulation. In case of multi-speaker TTS systems, voices with different characteristics is preferable, since it gives the user a wide spectrum of choices.

\begin{table}[h]
\begin{center}
\begin{minipage}{\textwidth}
\caption{Details of speakers}\label{tab1}%
\begin{tabular}{@{}lllll@{}}
\toprule
Speaker & Speaker ID  & Profession & Gender &Age\\
\midrule
Joji V. T.   & M1   & Linguist  & Male &28 \\
Sonia Jose Gomez     & F1   & TV Newsreader  & Female &43 \\
Jijo Joy   & M2   & Theatre Artist  & Male &26 \\
Greeshma S.   & F2  & Linguist  & Female  &22\\
Anil Arjunan    & M3   & Social Activist  & Male &48 \\
Vidya S.   & F3   & Theatre Artist  & Female &23 \\
Sonu S. Pappachan   & M4   & Political Science Researcher  & Male &25 \\
Simla Mujeeb Rahiman    & F4   & Research Assistant  & Female &24 \\
\botrule
\end{tabular}

\end{minipage}
\end{center}
\end{table}

\subsection{Recording of speech}

The recording was done in a soundproof recording studio in the Phonetics Lab of the Department of Linguistics, University of Kerala. A RODE microphone was used as the recording device, and the recording was sampled at 48 kHz, mono, with 24 bit depth. The equipment settings were tailored to studio conditions and the artist in question in order to produce high quality audio. The distance between the speaker and the microphone was set in the range of 4-6 inches.

The scheme for quality control was formulated after experimentation with the recording setup and recording process. Speaker F4, who was also a part of the project team, was engaged for experimentation in this regard. It was found that close monitoring and support during recording sessions is required to ensure the consistency of the read speech and to reduce error rate as much as possible. A project assistant closely monitored the recordings and noted down discrepancies in comparison with the text, if any. The voice quality was observed and intermittent breaks were given to the speakers to avoid fatigue. Multiple sentences were read in one stretch which was later segmented. Mismatches that occurred were corrected by rerecording the sentence. To ensure quality recordings, for certain difficult sentences, a trial reading was carried out before the actual recording. Particularly difficult sentences were discarded by each speaker when deemed necessary. 

\subsection{Post processing and speech corpus compilation}\label{postpro}

The recorded voice files were segmented into sentences automatically by detecting the silence between sentences. The automatically segmented speech was evaluated and corrected manually. Each audio file was examined in comparison with the corresponding text and corrections were made wherever required. All the sentences in the text corpus were uniquely labelled and the audio files carrying the corresponding speech were saved with a file name matching the label. 

The compiled speech corpus for each of the artists were again vetted by language experts for mismatches between the articulated phonemes in the audio file and those in the corresponding text file. Sentences with substantial mismatches between text and audio were discarded. If the mismatch was slight, then the text was corrected to match the audio. Through this process it was
ensured that the audio files and the corresponding text matched up to the
phonemic level. The resulting audio files were downsampled and saved as 16 kHz, 16-bit single-channel WAV files.

\section{Evaluation of the database using a deep learning TTS system}
Traditional neural TTS architectures use two separate components for generative modeling, splitting the process into two stages. The first stage generates from text an intermediate representation of speech features, such as mel-spectrograms, using an acoustic model. The second stage synthesizes a raw waveform from the intermediate representation via a neural vocoder. Those models are trained separately and then joined for inference. Two-stage pipelines, however, require a sequential and costlier training procedure, and their dependence on predefined intermediate features prevents applying learned hidden representations to improve performance further \citep{tan2021survey, kim2021conditional}.

The VITS network is a parallel end-to-end architecture for TTS that outperforms traditional two-stage architectures, and synthesizes natural sounding speech extremely close to human quality \citep{kim2021conditional}. VITS circumvents the issues laden in two-stage pipelines by connecting the two modules of TTS systems through latent variables to enable efficient end-to-end learning. It is used as a baseline comparison model for advancements in recent TTS methods and applications. Rijn et al. used a modified version of VITS for personalized voice generation \citep{van2022voiceme}, while Song et al. used multi-speaker VITS as a base architecture for talking face generation \citep{song2022talking}. Casanova et al. modified the network to achieve state-of-the-art results in zero-shot multi-speaker TTS and zero-shot voice conversion \citep{casanova2022yourtts}. Russell et al. trained VITS models for evaluating BU-TTS, a bilingual Welsh-English speech corpus \citep{russell2022bu}. 

Due to its simplified training procedure and improved performance, we decided to use VITS to evaluate the dataset over older but more common architectures for TTS systems such as Tacotron or FastSpeech. The model was separately trained for each of the 8 databases. The trained models were then used for inference to generate synthetic audio to be evaluated via a mean opinion score survey.


\begin{figure}[h]
\centering
\includegraphics[width=\textwidth]{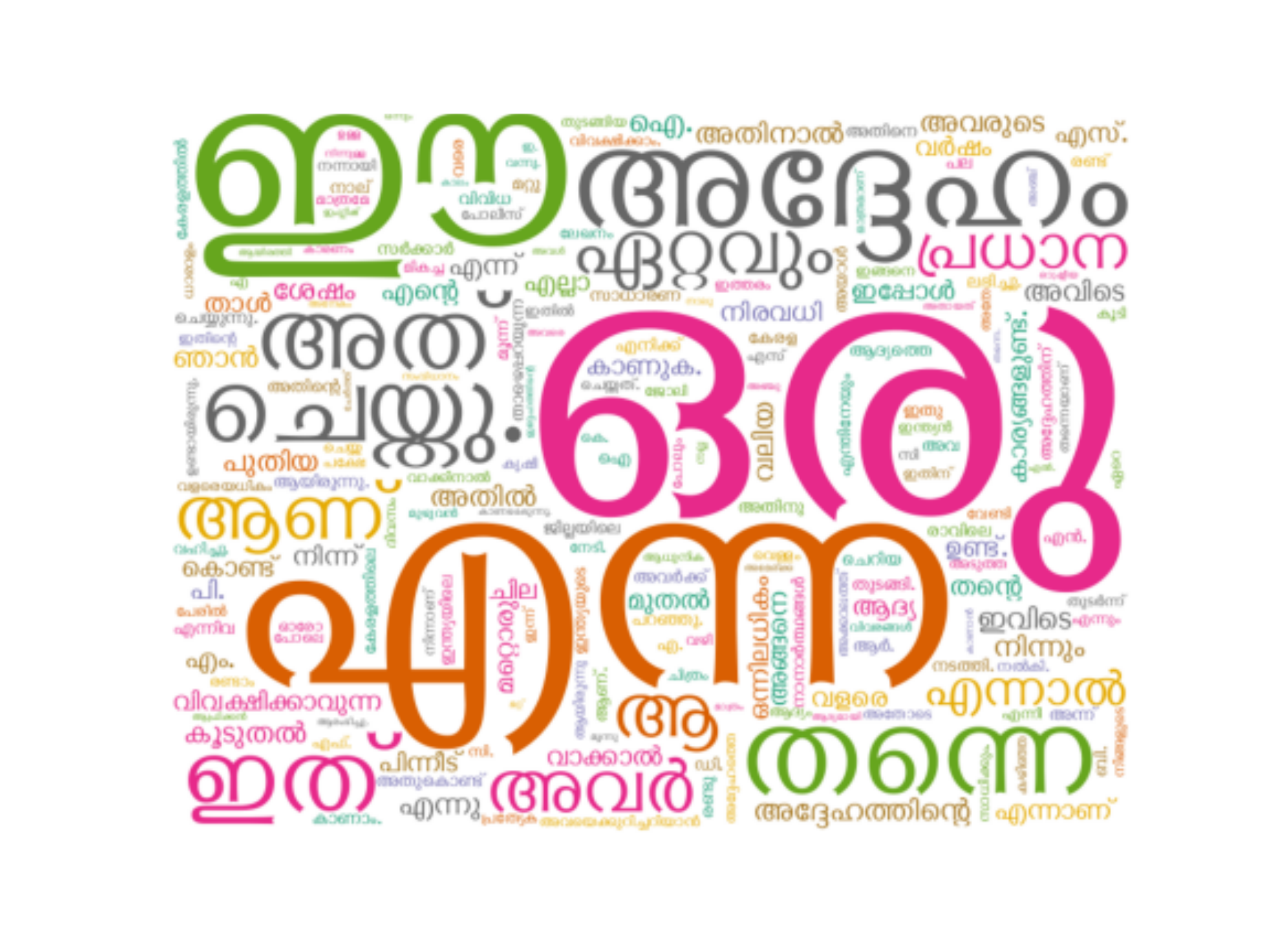}
\caption{Word cloud representation of the text in \emph{IMaSC}}\label{wc}
\end{figure}

\section{Results}\label{sec2}

\subsection{Details of \emph{IMaSC}}


\begin{table}[h]
\begin{center}
\begin{minipage}{0.9\textwidth}
\caption{Details of Speech corpus and corresponding text}\label{speechcorpus}%
\begin{tabular}{@{}lllllll@{}}
\toprule
\multirow{2}{*}{Speaker} & \multirow{2}{*}{Time (HH:MM:SS)} & \multirow{2}{*}{Sentences} & 
\multicolumn{2}{c}{Words} & \multicolumn{2}{c}{Phonemes} \\ \cmidrule(l){4-5} \cmidrule(l){6-7}
&&& Total & Unique & Total & Unique \\ \midrule
 M1  & 06:08:55 & 4,332  & 28,508  &15,912 & 239,066 & 56\\
 F1  & 05:22:39 & 4,294  & 28,196 &16,221 & 237,405 & 56\\
 M2  & 05:34:05 & 4,093  & 26,742 &15,223 & 226,715 & 56\\
 F2  & 06:32:39 & 4,416  & 29,015 &16,358 & 243,611 & 56\\
 M3  & 05:58:34 & 4,239  & 27,777 &15,937 & 235,163 & 56\\
 F3  & 04:21:56 & 3,242  & 21,489 &13,087 & 177,120 & 56\\
 M4  & 06:04:43 & 4,219  & 27,390 &15,599 & 233,467 & 56\\
 F4  & 09:34:21 & 5,638  & 36,664 &20,649 & 318,841 & 56\\
 \textbf{Total} & \textbf{49:37:54} & \textbf{34,473} & \textbf{225,781} & \textbf{23,604} & \textbf{1,911,388} & \textbf{56}\\
\botrule
\end{tabular}
\end{minipage}
\end{center}
\end{table}

\begin{figure}[h]
\centering
\includegraphics[width=\textwidth]{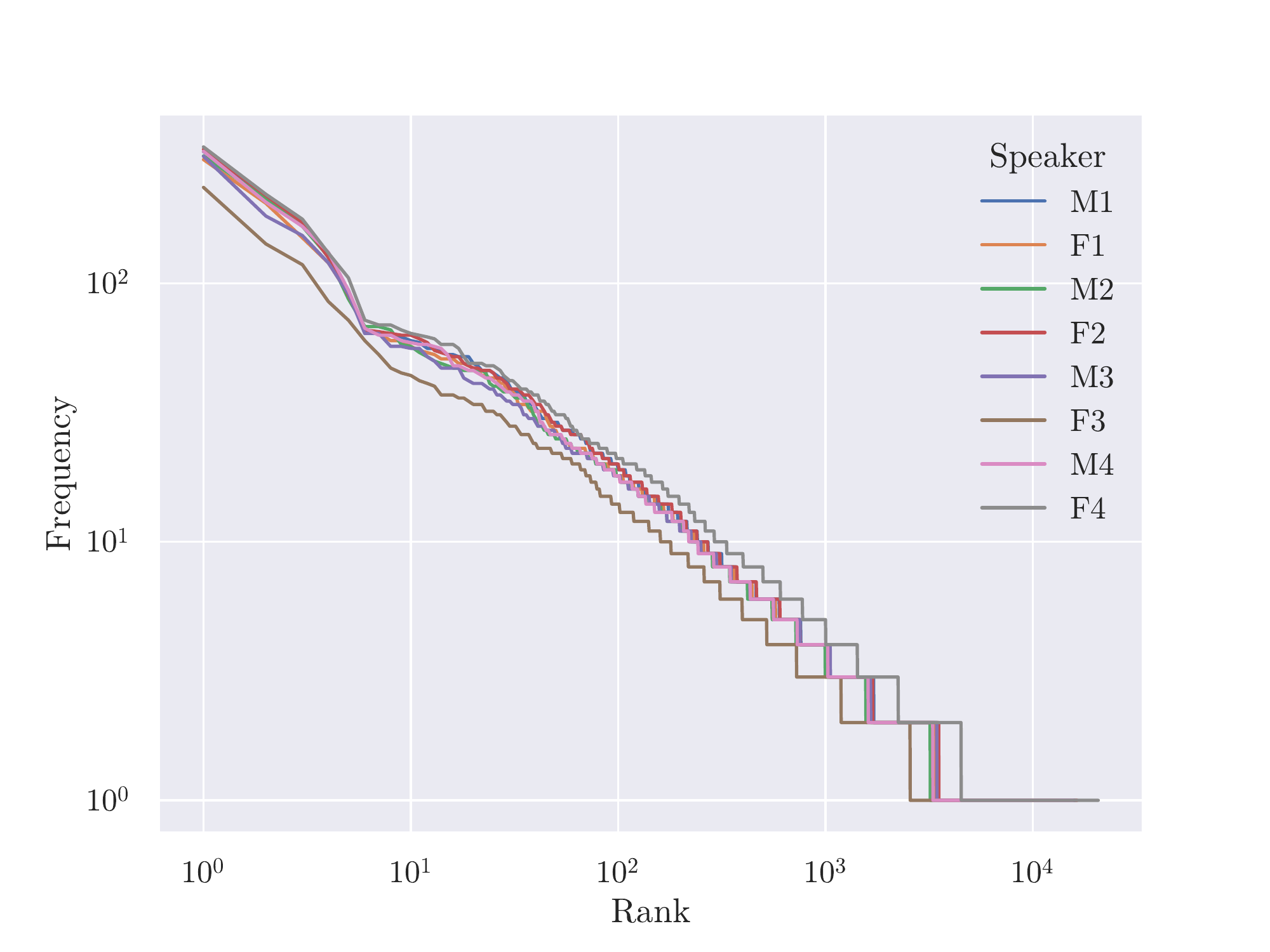}
\caption{Logarithmic plot of word rank versus word frequency in the text corpus of 8 speakers}\label{zipfwords}
\end{figure} 

\begin{figure}[h]
\centering
\includegraphics[width=\textwidth]{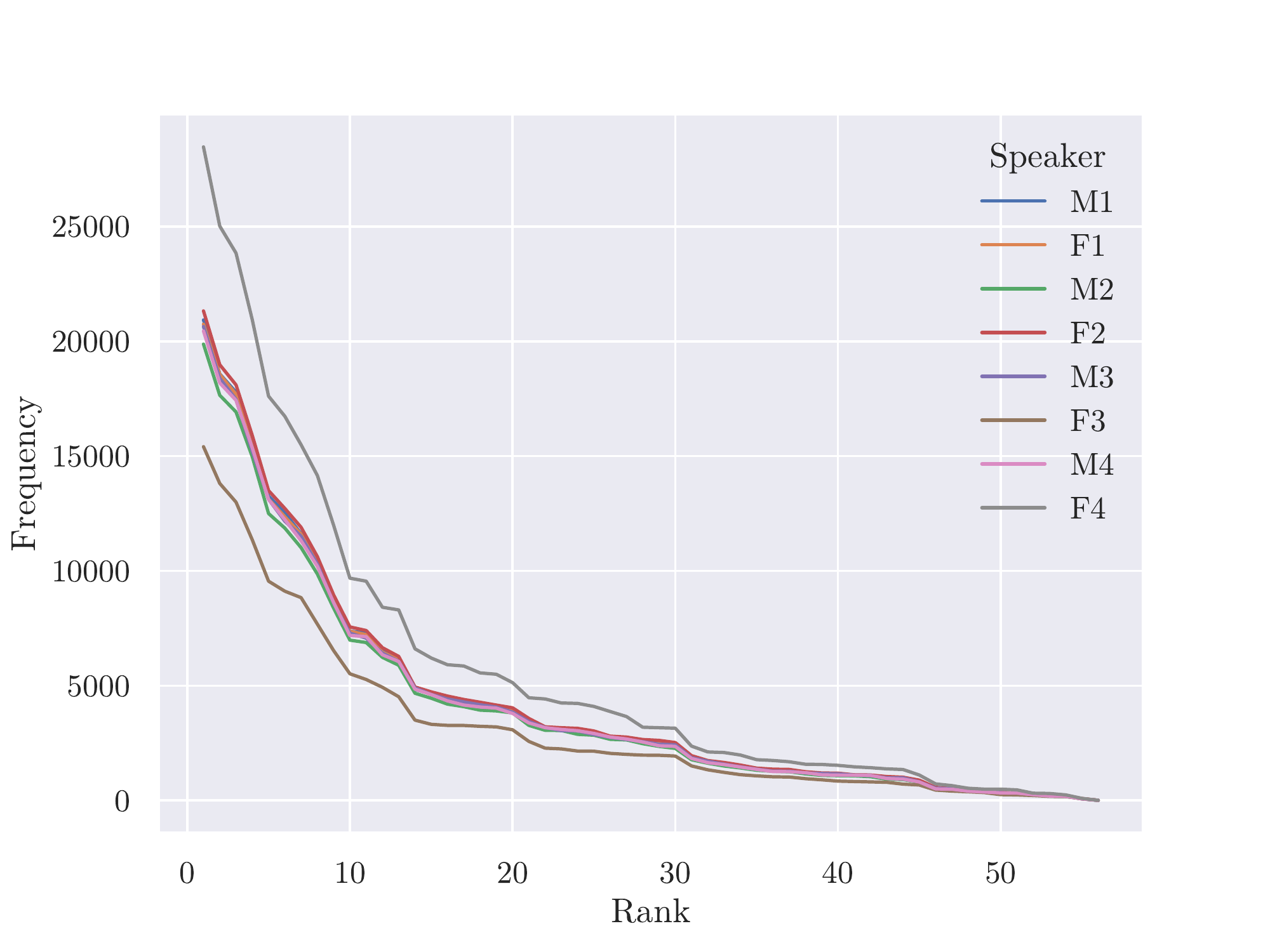}
\caption{Plot of phoneme rank versus phoneme frequency in the text corpus of 8 speakers}\label{zipfpho}
\end{figure} 

\begin{figure}[h]
\centering
\includegraphics[width=\textwidth]{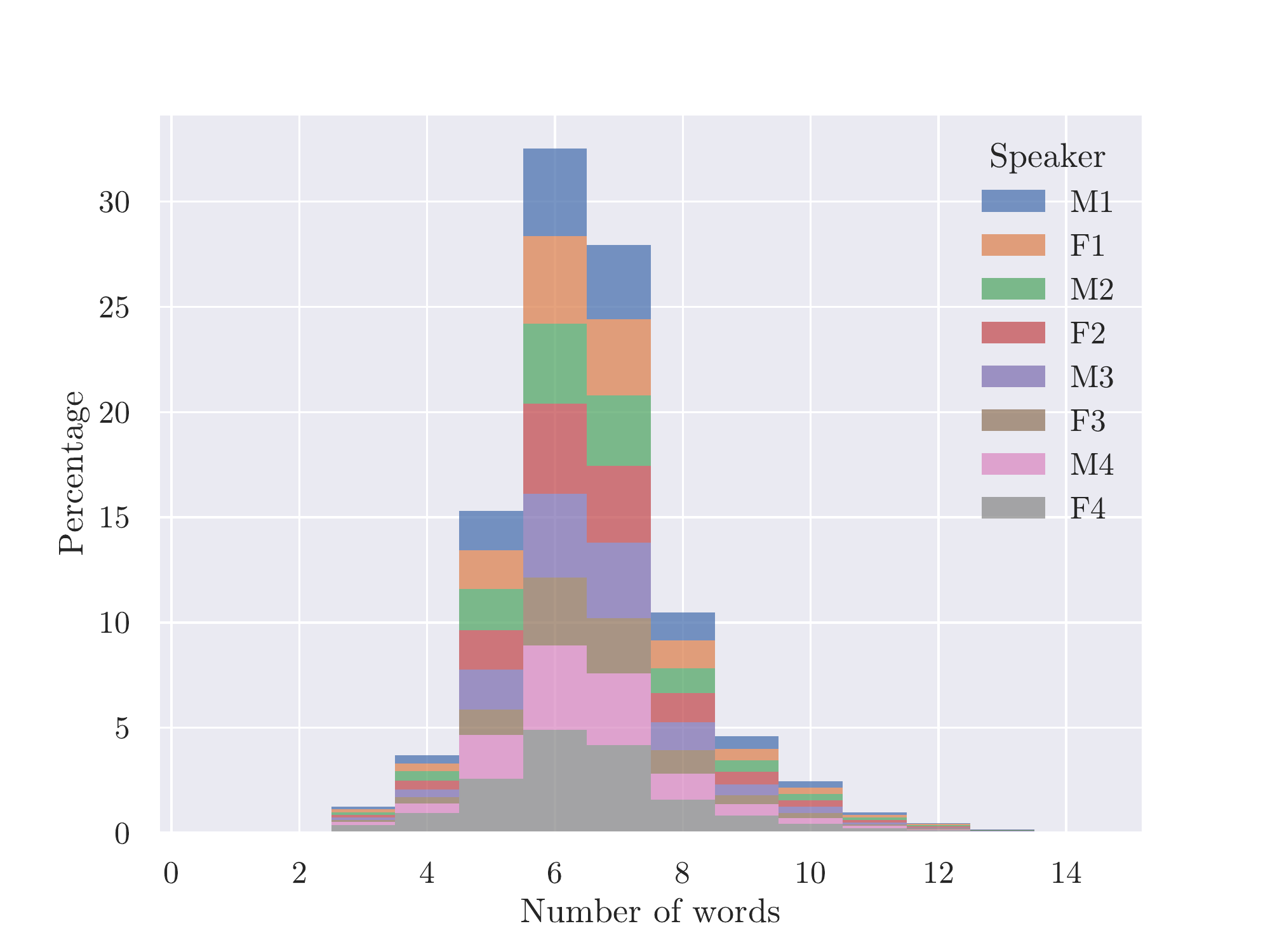}
\caption{Histogram of the number of words in each sentence}\label{histo}
\end{figure} 

\begin{figure}[h]
\centering
\includegraphics[width=\textwidth]{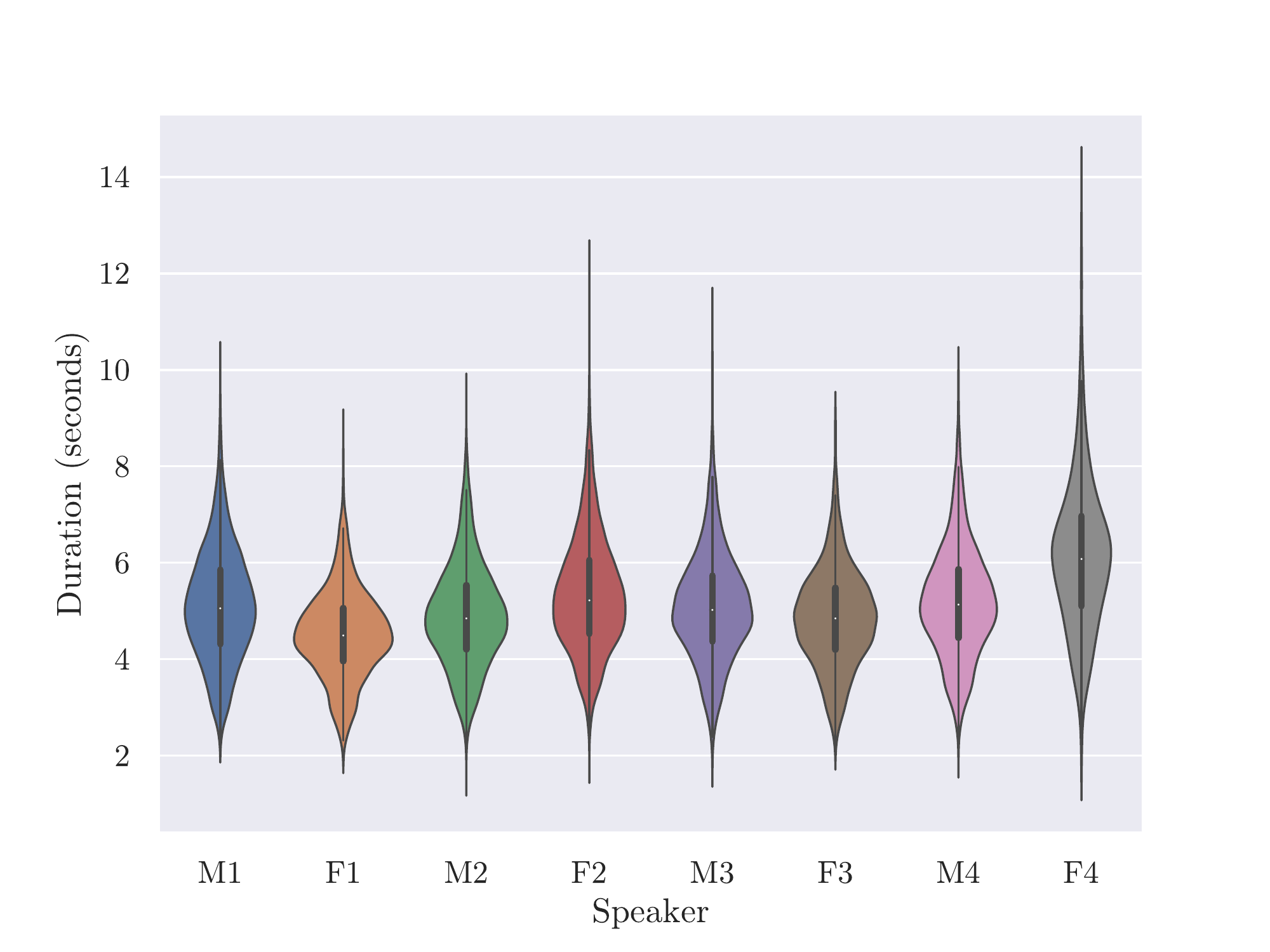}
\caption{Violin plot indicating the distribution of duration of audio samples for each speaker}\label{violin}
\end{figure}

488,249 sentences were obtained from Wikipedia dump, from which 8,853 unique sentences were obtained after linguistic evaluation and processing as outlined in sections \ref{corpus} and \ref{postpro}. The details of the speech corpus and the corresponding text of each speaker after quality check and data cleaning is given in Table \ref{speechcorpus}. The recorded speech of F4 is significantly higher than the rest of the speakers, since she was engaged in experimental recording sessions and her recorded speech was analysed for formulating the quality control process. 

The word cloud representation of the 36,589 words in the text corpus of \emph{IMaSC} is given in Fig. \ref{wc}. The word cloud indicates the most frequent words and provides a broad visual description of the words in the database. A close look into the word cloud reveals that the number of characters per word varies widely, indicating the agglutinative nature of Malayalam.

The frequency of the words in the text corpus for each speaker and its corresponding rank was found. The logarithmic plot of the word rank versus  frequency is given in  Fig. \ref{zipfwords}. It can be observed that the plot follows Zipf's law, which states that the frequency of any word is inversely proportional to its rank \citep{sicilia2002extension}. A plot of phoneme rank versus phoneme frequency, given in Fig. \ref{zipfpho}, also follows a power law. 

A histogram of the number of words per sentence read by the 8 speakers is given in Fig. \ref{histo}. It shows that more than 30\% of the sentences are 6 words long. It can be noted that F4 has spoken the longest sentences in terms of number of words (14 words).

The distribution of duration of audio samples for each speaker is given in Fig. \ref{violin}. It can be seen that the duration of the audio samples in general lie between 2s to 8s. The minimum variation is for speaker F1 (approximately 2s to 9s) and maximum for F4 (approximately 1s to 15s). Longer duration audio is to be expected for F4 as she has spoken comparatively longer sentences. The variation of distribution of duration between the speakers is due to diversity in the manner of articulation and variation in the set of sentences read by the speakers. 

\begin{figure}[h]
\centering
\includegraphics[width=\textwidth]{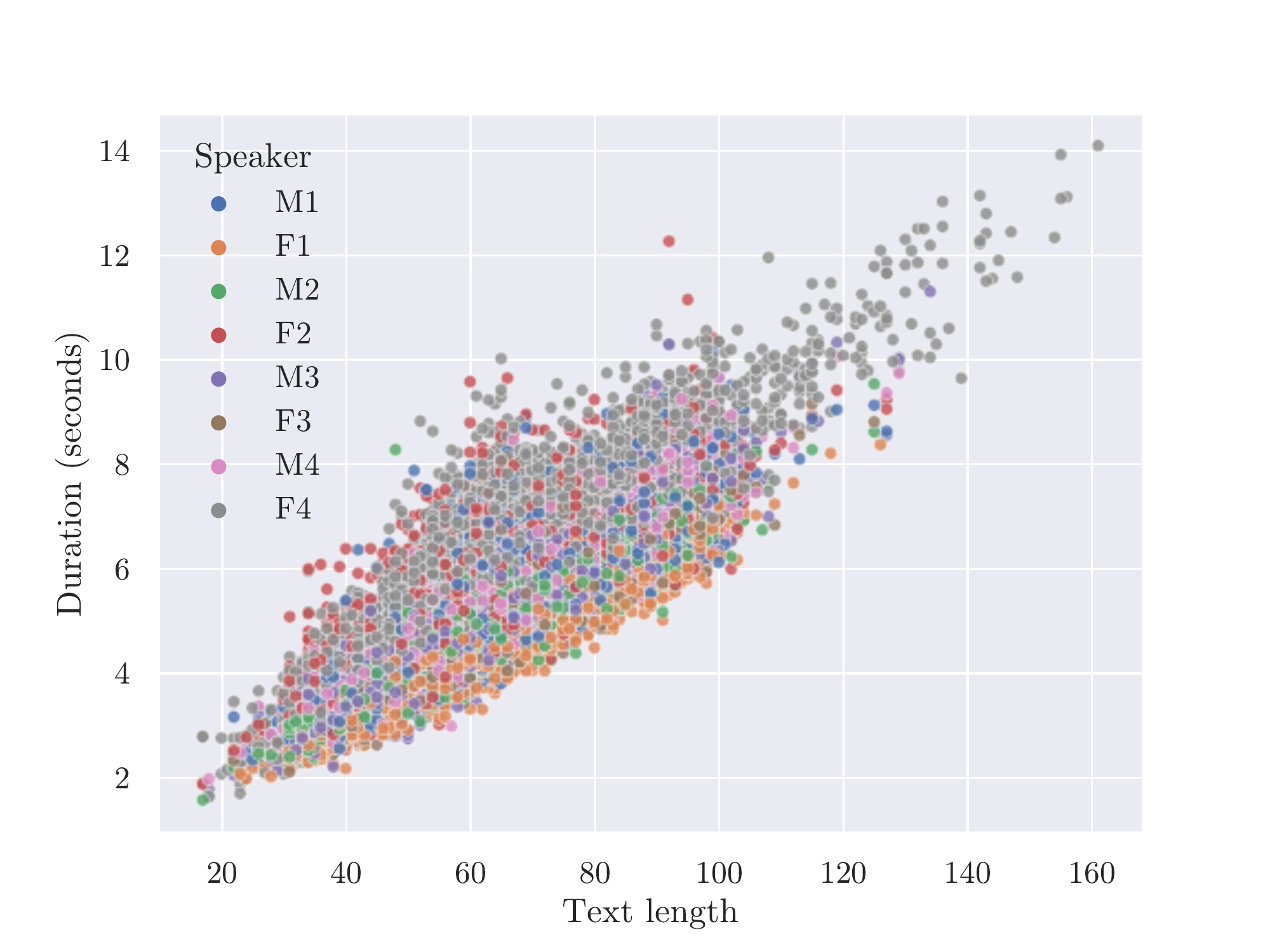}
\caption{Scatterplot of number of characters in each sentence and the duration of its corresponding audio}\label{scattlen}
\end{figure}

The scatterplot of the text length (number of characters) in each sentence and the duration of its corresponding audio is given in Fig. \ref{scattlen}. The text length for each sentence correlates linearly with corresponding audio duration as per the scatterplot. It also shows that text length mostly ranges between 40 to 100 and the corresponding duration between 2s to 8s. F4 has the audio sample with the maximum text length (160). The range of duration values as seen in this plot is in tune with that of the violin plot in Fig. \ref{violin}. 

\begin{figure}[h]
\centering
\includegraphics[width=\textwidth]{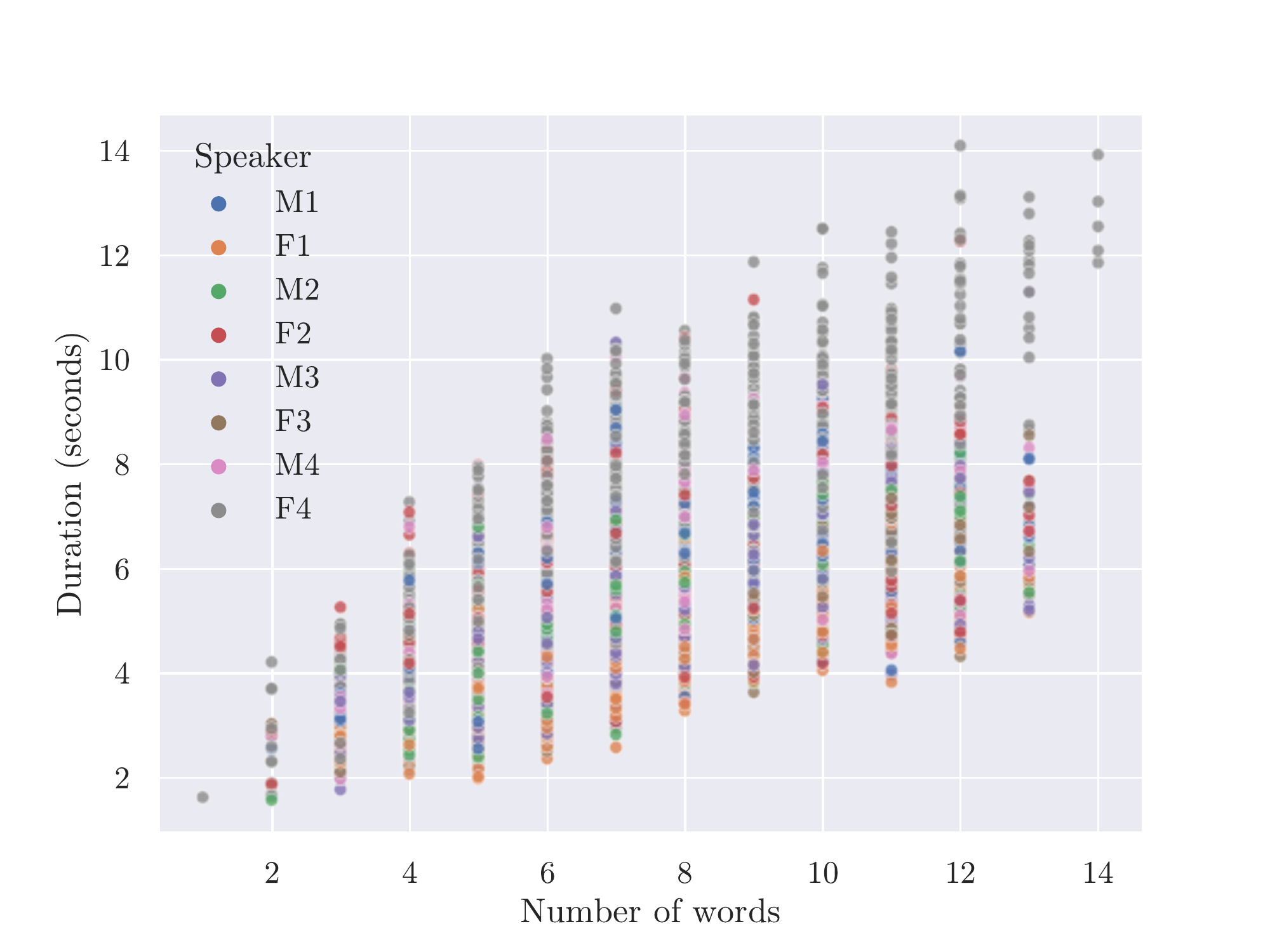}
\caption{Scatterplot of number of words in each sentence and the duration of its corresponding audio}\label{scattnum}
\end{figure} 

The number of words in the sentence also correlates similarly with audio duration as per the scatterplot in Fig. \ref{scattnum}, though with significantly more variation. This is again a consequence of agglutination in Malayalam, with the length of a word varying widely. For example, in the case of sentences with 6 words, the duration varies from 2s to 10s.  It is noted that F4 has more instances of longer sentences in terms of duration and number of words. The same fact is also seen in Fig. \ref{violin} and Fig. \ref{scattlen}.

The analysis of the corpus shows that \emph{IMaSC} has covered a large variety of words and consists of speech with different articulation styles. These features can make it a suitable candidate for training a deep learning TTS system with multiple speakers and speaking styles.

\subsection{Evaluation of \emph{IMaSC} using VITS}

We used the public VITS implementation available at Coqui TTS\footnote{\url{https://github.com/coqui-ai/TTS}}. The character set for Malayalam, including punctuation, was directly tokenized as Malayalam characters have close to a one-to-one correspondence with phonemes. Tokenized raw text and the corresponding speech were used for training. A separate model for each speaker was trained to evaluate their datasets individually. Each model was trained for 60k steps on a Tesla P100 GPU.

In TTS, the input character sequences and the output audio sequences are of different length, and VITS uses duration modeling and encoder-decoder attention mechanisms to capture the duration of influence of each character. In order to transform a sequence of characters $c$ to a larger sequence  of latent variables representing speech $z$, an alignment between the two sequences are required. In VITS, the alignment is a hard monotonic attention matrix with $\vert c \vert \times \vert z \vert$ dimensions representing how long each input character expands to be time-aligned with the target speech. 

\begin{figure}[p]
\centering
\includegraphics[width=\textwidth]{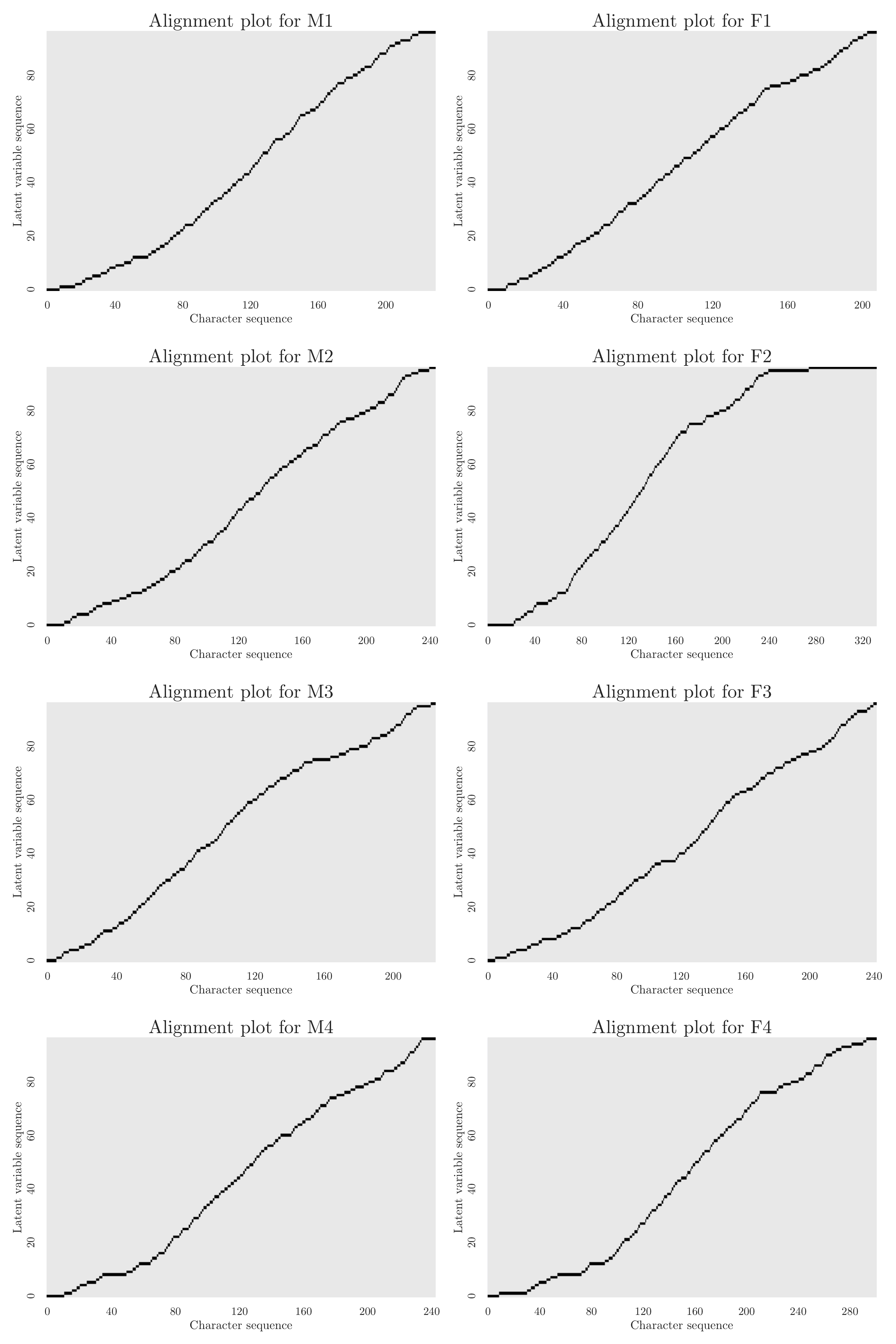}
\caption{The attention matrix visualized as a binary heatmap.}\label{attn}
\end{figure}

The visualized attention matrices of the trained models for synthesis of a sample sentence from the test set is given in Fig. \ref{attn}. We observe that the attention matrices look approximately diagonal, indicating that the sequence of latent variables is properly aligned with the sequence of input characters. 

\subsubsection{Mean Opinion Score}

We conducted a crowd-sourced Mean Opinion Score (MOS) test  with 20 participants for evaluating the models. 10 sentences were randomly selected from the test dataset to be synthesized, and 2 sentences, along with their corresponding audio, were chosen for evaluating ground truth. In the survey, each speaker thus had 12 text-audio pairs to be evaluated, repeated for all 8 speakers for a total of 96 questions. The different audio samples were each scored on a 5-point scale for naturalness, with 5 being excellent, 4 being good, 3 being fair, 2 being poor, and 1 being bad. 

Table \ref{MOS} gives the score obtained for each of the 8 speakers. The MOS for ground truth and synthesized speech for each speaker is detailed in the table. A comparison between our models and previous TTS systems for Malayalam is given in Table \ref{comparemos}. 

\begin{table}[h]
\begin{center}
\begin{minipage}{0.8\textwidth}
\caption{Mean Opinion Score of the speech synthesized from recorded database of each of the 8 speakers and average MOS}\label{MOS}%
\begin{tabular*}{\textwidth}{@{\extracolsep{\fill}}lll@{}}
\toprule
Speaker  & Synthesized Speech & Ground Truth \\

\midrule
 M1   & $4.60 \pm 0.09$ & $4.75 \pm 0.19$ \\
 F1   & $4.48 \pm 0.11$ & $4.75 \pm 0.19$ \\
 M2   & $4.38 \pm 0.11$ & $4.58 \pm 0.23$ \\
 F2   & $4.55 \pm 0.10$ & $4.65 \pm 0.26$ \\
 M3   & $4.46 \pm 0.12$ & $4.68 \pm 0.23$ \\
 F3   & $4.58 \pm 0.10$ & $4.90 \pm 0.14$ \\
 M4   & $4.46 \pm 0.12$ & $4.55 \pm 0.26$ \\
 F4   & $4.54 \pm 0.11$ & $4.58 \pm 0.23$ \\
\textbf{Average MOS}  & $\pmb{4.50} \pm \pmb{0.04}$ & $\pmb{4.68} \pm \pmb{0.08}$ \\
\botrule
\end{tabular*}

\end{minipage}
\end{center}
\end{table}

\subsection{Comparison of \emph{IMaSC} with other TTS corpora}
Table \ref{comparesize} provides a comparison of \emph{IMaSC} with other publicly available Malayalam TTS corpora. We note that \emph{IMaSC} is significantly larger than previous corpora both in number of sentences and hours of recorded audio. IndicSpeech employs a single female speaker while Baby et al. employs one male and one female, as compared to our 4 male and 4 female speakers. He et al. uses crowdsourcing to record audio and thus has 42 speakers (18 male and 24 female), but with only 5.51 hours of total speech, each speaker has an average of less than 8 minutes of speech. A single-speaker TTS system is thus infeasible. We also note that in contrast to the crowdsourcing approach, our speaker selection is deliberate and intended to represent a range of prosody and articulation styles while maintaining clear and comprehensive speech, as detailed in Section \ref{speakersel}.

The aforementioned corpora each have TTS systems trained on them, and we compare their reported MOS scores to our own in Table \ref{comparemos}. Our TTS system performs significantly better, with an average MOS score extremely close to ground truth. We note that He et al. trains two different multi-speaker TTS models for male and female speakers, and we average their separately reported MOS scores and confidence intervals.

\begin{table}[h]
\begin{center}
\begin{minipage}{0.7\textwidth}
\caption{Comparison of \emph{IMaSC} with other TTS corpora in terms of text size}\label{comparesize}%
\begin{tabular}{@{}lllll@{}}
\toprule
Database  & Sentences & Total Words  & Hours  & Number \\
 & & & & of speakers \\
\midrule
 Baby et al. & 11,300 & 58,098 & 17.89 & 2 \\
 IndicSpeech & 19,954 & 109,245 & 29.1 & 1 \\
 He et al.   & 4,126  & 25,330 & 5.51 &  \textbf{42} \\
 \textbf{IMaSC}       &  \textbf{34,473} &  \textbf{225,781} &  \textbf{49.63} & 8  \\
 
\botrule
\end{tabular}

\end{minipage}
\end{center}
\end{table}

\begin{table}[h]
\begin{center}
\begin{minipage}{0.6\textwidth}
\caption{Comparison of \emph{IMaSC} with other TTS corpora in terms of average MOS score of TTS built on each of these}\label{comparemos}%
\begin{tabular*}{\textwidth}{@{\extracolsep{\fill}}ll@{}}
\toprule
Model & Mean Opinion Score\\
\midrule
 Public API & $3.32$ \\
 IndicSpeech & $3.87$ \\
 He et al. & $4.01 \pm 0.13$ \\
 \textbf{IMaSC} & $\pmb{4.50} \pm \pmb{0.04}$ \\
 
\botrule
\end{tabular*}

\end{minipage}
\end{center}
\end{table}
\section{Conclusion}\label{sec13}

In this work, we presented \emph{IMaSC}, a Malayalam text and speech corpora that aims to address the lack of publicly available data for TTS applications. With 49 hours and 37 minutes of audio, 8 speakers and 34,473 text-audio pairs, \emph{IMaSC} is larger than any other public Malayalam text and speech corpus by a wide margin, with proper care taken for quality control.  We trained end-to-end TTS models for each speaker based on the VITS architecture to evaluate the database, and conducted a subjective MOS survey with 20 participants. We reported an average MOS score of 4.50 for speech synthesized with our models, close to the ground truth of 4.68. With an average of more than 6 hours per speaker across 8 speakers, \emph{IMaSC} will enable the development of multi-speaker text to speech synthesis systems, which provides the user with a choice of selecting synthesized speech with prosody and other voice qualities of their choice.

\bmhead{Acknowledgments}

We would like to thank the International Centre for Free and Open Source Software (ICFOSS), an autonomous organization set up by the Government of Kerala for promoting free and open source software for funding the project. We also acknowledge the support rendered by Shijith S. and other linguists at Department of Linguistics, University of Kerala in different stages of the database creation.


\bibliography{sn-bibliography}

\end{document}